\documentstyle[11pt]{article}
\input amssymb.sty

\title{Quantum Superpositions Do Exist!\\
But `Quantum Physical Reality $\neq$ Actuality'\\(Reply to Dieks and Griffiths)}

\author{{\sc Christian de Ronde}\thanks{Fellow Researcher of the Consejo
Nacional de Investigaciones Cient\'{\i}ficas y T\'ecnicas.}}
\date{}

\begin{document}

\bibliographystyle{plain}
\maketitle

\begin{center}
\begin{small}
Philosophy Institute ``Dr. A. Korn" \\ 
Buenos Aires University, CONICET - Argentina \\
Center Leo Apostel and Foundations of  the Exact Sciences\\
Brussels Free University - Belgium \\
\end{small}
\end{center}

\begin{abstract}
\noindent In this paper we analyze the definition of quantum superpositions within orthodox Quantum Mechanics (QM) and their relation to physical reality. We will begin by discussing how the metaphysical presuppositions imposed by Bohr on the interpretation of QM have become not only interpretational dogmas which constrain the limits of the present Orthodox Line of Research (OLR), but also how these desiderata implicitly preclude the possibility of developing a physical representation of quantum superpositions. We will then continue analyzing how most interpretations of QM argue against the existence of superpositions. Firstly, we will focus on those interpretations which attempt to recover a classical representation about ``what there is'', and secondly, we will concentrate on the arguments provided by Dieks and Griffiths who, staying close to the orthodox formalism, also attempt to ``get rid of the ghost of Schr\"odinger's cat''. Contrary to the OLR, we will argue ---based on our definition of Meaningful Physical Statements (MPS)--- that from a representational realist perspective which stays close to the orthodox Hilbert space formalism, quantum superpositions are not only the key to the most important ---present and future--- technological and experimental developments in quantum information processing but also, they must be considered as the kernel of any interpretation of QM that attempts to provide a physical representation of reality. We will also argue that the price to pay for such representational realist development must be the abandonment of the (dogmatic) idea that `Actuality = Reality'.
\end{abstract}
\begin{small}


\end{small}

\newtheorem{theo}{Theorem}[section]

\newtheorem{definition}[theo]{Definition}

\newtheorem{lem}[theo]{Lemma}

\newtheorem{met}[theo]{Method}

\newtheorem{prop}[theo]{Proposition}

\newtheorem{coro}[theo]{Corollary}

\newtheorem{exam}[theo]{Example}

\newtheorem{rema}[theo]{Remark}{\hspace*{4mm}}

\newtheorem{example}[theo]{Example}

\newcommand{\proof}{\noindent {\em Proof:\/}{\hspace*{4mm}}}

\newcommand{\qed}{\hfill$\Box$}

\newcommand{\ninv}{\mathord{\sim}} 

\newtheorem{postulate}[theo]{Postulate}

\vspace{0.5cm}

\begin{flushright}
\emph{When the layman says `reality' he usually thinks  
\\that he is speaking about something which is self-evidently known; \\
while to me it appears to be specifically\\ 
the most important and extremely difficult task of our time\\
to work on the elaboration of a new idea of reality.}\\
\vspace{0.5cm}
Wolfgang Pauli
\end{flushright}

\section*{Introduction}

Quantum superpositions are being used today in laboratories all around the world in order to develop the most outstanding technological and experimental developments of the last centuries. Indeed, quantum computation, quantum teleportation, quantum cryptography and the like technologies are opening an amazing range of possibilities for the near future. As it is well known, this new quantum technological era is based and founded on one of the main principles of Quantum Mechanics (QM), the so called {\bf superposition principle} ---which in turn gives rise to quantum superpositions and entanglement. But while many experimentalists state that Schr\"odinger's cats are getting ``fatter'' and ``bigger'', while it becomes more and more clear that there is something real about these quantum superpositions, philosophers of QM in charge of analyzing and interpreting these mathematical expressions (through the many interpretations of QM that can be found in the literature) coincide almost exclusively ---even though for very different reasons--- that quantum superpositions do not exist! It is indeed true that quantum superpositions do not seem to  match our picture of ``classical reality''; but even for those who seem to accept ---like Dieks and Griffiths--- that QM is not talking about ``classical reality'', there is a strange need of getting ``rid of the ghost of Schr\"odinger's cat'' at any cost. But why? Why is it so difficult to come up with a coherent interpretation of quantum superpositions that relates them to physical reality instead of mere measurement outcomes? We will argue in this paper that one of the main reasons is related to the role played by a deeply rooted metaphysical equation which states that: `Actuality = Reality'. We will argue that if we are willing to discuss the possibility that ``Quantum Physical Reality $\neq$ Actuality', then there is plenty of space to interpret and represent quantum superpositions in terms of (non-actual) physical reality. The paper is organized as follows. In section 1, we discuss what should be considered as a representational realist stance about physics which, going against ``naive realism'', accepts as a standpoint the theory-ladenness of physical experience. Section 2 provides an analysis of the metaphysical presuppositions imposed by Bohr on the possible interpretations of QM. In section 3, we discuss how such metaphysical presuppositions have become unquestionable dogmas which guide the Orthodox Line of Research (OLR), more specifically we analyze how two of the main problems in the foundational literature about QM, namely, the `measurement' and `basis' problems, focus only on the justification of the classical representation of physics. Section 4 provides a definition of MPS within a theory and discusses the need to consider Counterfactual Reasoning (CR) as a {\it necessary condition} for providing a coherent physical discourse and representation of reality. In section 5 we analyze and discuss, firstly, those interpretations that deny the existence of quantum superpositions ``right from the start'' by claiming that QM should be represented ---even changing the formalism--- in terms of ``classical reality'', and secondly, we will focus on those interpretations which even though agree that QM does not talk about ``classical reality'', still deny that quantum superpositions are related to physical reality. In particular,  we will concentrate on Dieks' and Griffiths' arguments against the existence of quantum superpositions. In section 6, we remark ---given that physics is an experimental science--- the relation of quantum superpositions to present experimental and technological developments in quantum information processing that are taking place today in the lab. In section 7, we continue providing a series of arguments in favor of considering quantum superpositions as real physical existents. We also stress the need of leaving aside the classical metaphysical presupposition that `Actuality = Reality'. In the final section we give the conclusions of the paper.

\section{The Representational Realist Stance}

Physics has been always connected to different philosophical stances, but certainly, realism is one of the main viewpoints within the history of physics. The main presupposition of realism with respect to physics is that physical theories talk about reality. According to our representational realist stance \cite{deRonde11, deRonde14a}, which will be presupposed through the rest of the paper, the fundament of any physical theory is physical representation and not experimental data ---the latter should be regarded by a representational realist only as part of the confirmation (or failure) of the empirical adequacy of a theory. Indeed, the representational realist takes as a standpoint the existence of Nature adding to it the idea that such existence can be represented through the interrelation of mathematical formalisms and conceptual networks allowing us to predict and understand specific phenomena. This realist stance, which relates to Heisenberg's closed theory approach \cite{Bokulich04}, goes against any type of naive realism that denies the theory ladenness of physical experience. Against the idea that one could distinguish between phenomena and raw observable data we have argued in \cite{deRonde14a} that even a `click' in a detector or a `spot' in a cloud chamber are only determined through the logical and ontological principles of existence, non-contradiction and identity. Such principles are not something that we find out in the world but rather metaphysical presuppositions that shape even our most basic experience. We have argued that even `clicks' and `spots' found in a laboratory are theory laden. Any stance going against representational realism must be capable of producing arguments that explain how physical experience can account for phenomena without the need of presupposing a physical representation.

Physical representation allows us to think about experience and predict phenomena without the need of actually performing any experiment. This is of course a completely different standpoint from those of many empiricist approaches who argue instead that the fundament of physics is `actual experimental data'. For example, as remarked by van Fraassen \cite[pp. 202-203]{VF80}: ``To develop an empiricist account of science is to depict it as involving a search for truth only about the empirical world, about what is actual and observable.'' Even though there are different positions with respect to the consideration of what accounts to be an empirically adequate theory (see e.g. \cite[p. 351]{BogenWoodward}) we would like to draw a line distinguishing between those stances that  accept the theory ladenness of physical experience and those which deny it. Our analysis is only concerned with the former.\footnote{We acknowledge however that the theory ladenness of physical experience is still today controversial within philosophy of science. As remarked by Bogen and Woodward \cite[p. 304]{BogenWoodward}: ``The positivist picture of the structure of scientific theories is now widely rejected. But the underlying idea that scientific theories are primarily designed to predict and explain claims about what we observe remains enormously influential, even among the sharpest critics of positivism.''}

Following van Fraassen \cite[p. xviii]{VF02}, we have called the attention to the importance of making explicit the metaphysical stance that one takes in order to analyze a specific problem \cite{deRonde11, deRonde14a}. In this respect we would like to make clear right from the start what should be considered to be the kernel of a representational realist account of physics:\\

\noindent {\it {\bf Representational Realism about Physical Theories (RRPT):} A representational realist account of a physical theory must be capable of providing a physical representation of reality in terms of a network of concepts which relates to the mathematical formalism of the theory and allows to make predictions of definite phenomena.}\\ 

\noindent Contrary to our definition of realism in physics which considers representation as a main construct of physical theories, naive realism claims instead that physical observation provides direct access to reality {\it as it is}. This idea was already implicit in the logical positivist distinction between {\it theoretical terms} and {\it empirical terms}. But even though in the philosophy of science community this distinction is characterized as ``naive'', many of the problems discussed in the literature still presuppose implicitly such distinction. Indeed, as remarked by Curd and Cover \cite[p. 1228]{PS}: ``Logical positivism is dead and logical empiricism is no longer an avowed school of philosophical thought. But despite our historical and philosophical distance from logical positivism and empiricism, their influence can be felt. An important part of their legacy is observational-theoretical distinction itself, which continues to play a central role in debates about scientific realism.'' We have argued that the naive realist stance is not only philosophically untenable but maybe even more importantly, closes the door to a fundamental development of physics ---since such stance assumes we already know what reality {\it is} in terms of (naive) observation.

From a realist perspective, physics attempts to describe a world in which we humans have no special preeminence with respect to existence. In this respect, the description or representation provided by classical physics was clearly specified since Newton's mechanics in terms of systems constituted by definite valued properties; i.e., in general terms, what is called an {\it Actual State of Affairs} (ASA).\footnote{See for discussion and definition of this notion in the context of classical physics \cite{RFD14}.} Also from a realist viewpoint, measurement and observation have been always considered as a way of exposing or discovering such preexistent ASA. But, as we know ---contrary to classical physics--- QM places serious difficulties for such a realist representation. An evidence of the deep crisis of physical representation within the theory of the quanta is the fact that more than one century after its creation the physics community has reached no consensus about what the theory is talking about. Indeed, for many a consistent interpretation that would match this strange formalism to a physical representation of reality seems to difficult to be found, for others it is enough for QM to account for the correct measurement outcomes. 

Of course, when discussing about QM and its interpretation there are multiple standpoints and interpretative strategies that one can assume. For example, one can argue ---as it has been done already by Fuchs and Peres \cite[p. 70]{FuchsPeres00}--- that ``[...] quantum theory does not describe physical reality. What it does is provide an algorithm for computing probabilities for the macroscopic events (``detector clicks") that are the consequences of experimental interventions. This strict definition of the scope of quantum theory is the only interpretation ever needed, whether by experimenters or theorists." This instrumentalist perspective is satisfied with having an algorithmic recipe that allows us to calculate measurement outcomes from the formalism. But this is certainly not enough for a representational realist, for whom the formalism should be capable of relating coherently an interpretation which allows to provide a physical representation of reality according to the theory \cite{deRonde14a}. In contrast, within a representational realist approach to QM there are two main possibilities. The first one is to argue that QM makes reference to the same physical representation provided by classical physics; i.e. that it talks about an ASA. This is, for example, the main idea presupposed by the Hidden Variable Program (HVP) which, as noticed by Bacciagaluppi  \cite[p. 74]{Bacciagaluppi96}, attempts to ``restore a classical way of thinking about {\it what there is}.'' The second possibility is to consider that QM might describe physical reality in a different, maybe even incommensurable, way to that of classical physics. This possibility seems to be endorsed by Griffiths \cite[p. 174]{Griffiths13} who argues that many of the problems with the interpretation of QM come from ``the view that the real world is classical, contrary to all we have learned from the development of quantum mechanics in the twentieth century.'' However, this second possibility, of developing a non-classical representation of physical reality has not in our opinion been thoroughly investigated. The reason is that the metaphysical presuppositions imposed by Bohr to the OLR implicitly take as a standpoint the idea that `Actuality = Reality' denying the possibility of a non-classical realm of reality.

\section{Niels Bohr's Creation of Orthodox Dogmas}

In \cite{deRonde15a, deRonde15b}, we have argued extensively that Bohr is the main responsible for producing an epistemological interpretation of QM that does not only limit physical representation in terms of classical language and classical phenomena but also precludes the very possibility of introducing and developing new (non-classical) concepts. This was done by Bohr through the introduction of two main desiderata which will be discussed in the following. Unfortunately, since the mid 20th century the OLR took these desiderata as necessary standpoints to think about the interpretation of QM. Accepted as unquestionable, these metaphysical presuppositions were turned into dogmas that any interpretation of QM had to respect.

The first metaphysical presupposition is the idea that there must exist a ``quantum to classical limit'', implying what Bokulich calls an ``open theory approach'' \cite{Bokulich04}. This idea was put forward by Bohr in terms of his {\it correspondence principle} \cite{BokulichCP}.

\begin{enumerate}
\item {\bf Quantum to Classical Limit:}  The principle that one must find a ``bridge'' or ``limit'' between classical mechanics and QM.
\end{enumerate}

The second metaphysical principle which has guided the OLR can be also traced back to Bohr's claim that physical experience needs to be expressed exclusively in terms of classical physical language \cite{BokulichBokulich}. Bohr \cite[p. 7]{WZ} stated that: ``[...] the unambiguous interpretation  of any measurement must be essentially framed in terms of classical physical theories, and we may say that in this sense the language of Newton and Maxwell will remain the language of physicists for all time.'' In this respect [{\it Op. cit.}, p. 7], ``it would be a misconception to believe that the difficulties of the atomic theory may be evaded by eventually replacing the concepts of classical physics by new conceptual forms.'' 

\begin{enumerate}
\item[2.] {\bf Classical Representation of Physics:} The principle that one needs to presuppose the classical representation of physics in order to discuss about {\it phenomena} and interpret QM.
\end{enumerate}

\noindent Both principles go clearly against a radical non-classical understanding of QM. A direct consequence of such commitment against new non-classical notions has been the incoherent analysis of the double-slit experiment which, described in terms of QM, cannot be understood as talking neither about `waves' nor `particles'. As a matter of fact we know quite well that  $\Psi$ cannot be interpreted in terms of waves or particles for many reasons that were collected since the early discussions of the founding fathers. Let us shortly recall some of them. Firstly, $\Psi$ is a mathematical entity that lives in configuration space, not in classical 3-dimensional space ---which in turn would allow an interpretation in terms of the Newtonian physical notions of space and time. Consequently, $\Psi$ cannot represent a `particle' nor a `wave' which are physical notions that live in a classical 3-dimensional space. Secondly, according to the orthodox Born interpretation of $\Psi$, the quantum wave function describes a probability distribution ---something difficult to imagine since the quantum probability is non-Kolmogorovian and thus cannot be interpreted in terms of ignorance about an ASA \cite{Redei12}--- and thus it is neither a `wave' nor a `particle'. Thirdly, the fact that the phenomena is wave-like (or particle-like) does not necessarily imply that one is talking about waves (or particles) for in a physical theory one also needs to be capable of relating the phenomena not only to physical notions but also to a mathematical formalism which is coherent with such notions. Fourthly, a `click' in a typical quantum experimental set up does not behave as if that which is making the click is a `particle' or a `wave'. Bell's inequalities prove explicitly that `quantum clicks' cannot be represented in terms of a classical local-realistic theory ---to which `particles' and `waves' pertain. There are well known experiments that have tested the weirdness of such non-classical `clicks'. Fifthly, the Kochen-Specker (KS) theorem \cite{KS} proves that $\Psi$ does not possess definite valued properties, in contrast with the case of `waves' and `particles' which do possess definite valued properties. Finally, a recent theorem has proven that $\Psi$ cannot be represented in terms of dynamical properties \cite{deRondeMassri14}. 

It is quite clear that the knowledge we have acquired of QM today is more detailed and accurate than the one Bohr and Einstein had at the beginning of the 20th-Century when they discussed the interpretation of QM in terms of the wave-particle duality. With the knowledge we have today it would also seem wise to recall the old logical positivist lesson that the use of inappropriate notions within language can only create pseudoproblems. Unfortunately, today, even within the specialized literature, it is extremely common to read arguments which still use uncritically the notions of `wave' and `particle' in order to analyze the interpretation of QM. Those who speak in this way, when confronted to some of the just mentioned arguments might go down a different path and claim that: ``This is just a way of speaking. Everybody knows that `quantum particles' are different of `classical particles'." Then of course the question of any interested interlocutor would be: ``But then, what is a quantum particle?" And then the answer is always very conclusive: ``Quantum particles are something really weird!"

\section{The (Orthodox) `Measurement' and `Basis' Problems Revisited}

As analyzed in \cite{deRonde15a} the OLR deals with a specific set of problems which analyze QM from a classical perspective. This means that all problems assume as a standpoint a classical representation and only reflect on the formalism in ``negative terms'', that is, in terms of the failure of QM to account for the classical representation of reality and its concepts: separability, space, time, locality, individuality, identity, actuality, etc. The ``negative'' problems are thus: {\it non-}separability, {\it non-}locality, {\it non-}individuality, {\it non-}identity, etc.\footnote{I am grateful to Bob Coecke for this linguistic insight.} These problems start their analysis from the notions of classical physics assuming implicitly as a standpoint the strong metaphysical presupposition that QM should be able to represent physical reality according to such classical notions. But among the many problems that can be found in the literature there are two unsolved main problems which show most explicitly that accommodating QM within classical physics seems an impossible task. The first problem relates directly to the issue of contextuality and is called the ``basis problem'':\\ 

\noindent {\it {\bf Basis Problem (BP):} Given the fact that $\Psi$ can be expressed by multiple incompatible bases ---given by the choice of a Complete Set of Commuting Observables (CSCO)--- and that due to the KS theorem the observables arising from such bases cannot be interpreted as simultaneously preexistent, the question is: how does Nature make a choice between the different bases? Which is the objective physical process that leads to a particular basis instead of a different one?}\\

\noindent Once again, the BP is a way of discussing quantum contextuality in ``negative terms''. The problem already sets the solution through the specificity of its questioning. The problem presupposes that there is a path ---in accordance to the quantum to classical limit imposed by Bohr---  from the weird contextual quantum formalism to a classical noncontextual experimental set up in which classical discourse holds. If one could explain this path through an objective physical process then the choice of the experimenter could be regarded also as part of an objective process as well ---and not one that determines reality. Unfortunately, still today the problem remains with no solution within the limits of the orthodox formalism. There is no physical representation of the process without the explicit change of the formalism of the theory or the addition of strange {\it ad hoc} rules; rules which not only lack any physical justification but, more importantly, also degrade the MPS of the theory (section 5). 

A very different problem ---sometimes also mixed and partly confused with the BP--- is the so called ``measurement problem'' which deals explicitly with the superposition principle and takes as a standpoint a specific basis or context.\\

\noindent {\it {\bf Measurement Problem (MP):} Given a specific basis (or CSCO) QM describes mathematically a state in terms of a superposition (of states), since the evolution described by QM allows us to predict that the quantum system will get entangled with the apparatus and thus its pointer positions will also become a superposition,\footnote{Given a quantum system represented by a superposition $\sum c_i | \alpha_i \rangle$, when in contact with an apparatus ready to measure, $|R_0 \rangle$, QM predicts that system and apparatus will become ``entangled'' in such a way that the final `system + apparatus' will be described by  $\sum c_i | \alpha_i \rangle  |R_i \rangle$. Thus, as a consequence of the quantum evolution, the pointers have also become ---like the original quantum system--- a superposition of pointers $\sum c_i |R_i \rangle$. This is why the MP can be stated as a problem only in the case the original quantum state is described by a superposition of more than one term.} the question is why do we observe a single outcome instead of a superposition of them?}\\

\noindent The measurement problem is also a way of discussing the formalism in ``negative terms'' with respect to classical physics. In this case the problem concentrates in the justification of measurement outcomes. It should be remarked that the MP presupposes that the basis (or context) ---directly related to a measurement set up--- has been already determined. Thus it should be clear that there is no question regarding the contextual character of the theory within this specific problem. But once the experimental arrangement is settled ---leaving aside the BP--- a new problem appears: due to the superposition principle one can find, within a context, that the state is mathematically described in terms of the so called ``quantum superpositions'' which are weird mathematical expressions composed by sets of states, $| \alpha_i  \rangle$, each of which gives rise to (compatible) projectors, $| \alpha_i \rangle \langle \alpha_i |$, interpreted as {\it compatible observables} (or properties). As we have discussed in \cite{daCostadeRonde13} such quantum superpositions can be, in general, composed of {\it contradictory properties} (section 7.3). Notice that given a $\Psi$ we call a quantum superposition to each different representation of the $\Psi$ in a specific basis. Thus the $\Psi$ gives rise to different superpositions, each one of them determined within a CSCO. This goes against the orthodox assumption that every superposition arising from $\Psi$ is ``the same'' superposition irrespectively of the basis, implying through this idea a reintroduction of contextuality within the MP. This interpretation confuses the whole problem and changes the issue at stake. We have extensively discussed this interpretational maneuver and the meaning of quantum contextuality in \cite{deRonde15c}. 

In the MP the weirdness appears because a superposition can be composed simultaneously by a specific property (e.g., the system has the property `spin up in the x direction') and its contradictory property (e.g., the system has the property `spin down in the x direction').\footnote{There is an ongoing debate regarding the interpretation of such properties in terms of `contradiction' and `contrariety'. See: \cite{ArenhartKrause14a, ArenhartKrause14b, ArenhartKrause14c, deRonde15a}.} On the one hand, due to the seeming violation of the Principle of Non-Contradiction (PNC), it makes no sense to interpret both properties as actual ones (see for discussion \cite{daCostadeRonde13}). This was cleverly exposed by Schr\"odinger in his 1935 {\it gedankenexperiment} in which a cat, after interacting with an atom represented by a quantum superposition, possessed at the same time the property of being `alive' and the property of being `dead' \cite{Schr35}. On the other hand, all terms in the superposition must be considered in the evolution of the state and the predictions that can be made from it. One could analyze what such weird quantum superpositions represent in physical terms, instead the orthodox literature ---following Bohr--- has limited the analysis through the MP to the justification of actual measurement outcomes.

\section{Meaningful Physical Statements and Counterfactual Reasoning}

In order to discuss and analyze physical interpretations of a theory one should first agree on what should be considered as {\it Meaningful Physical Statements} (MPS) within that theory. Furthermore, from a representational realist perspective, the theory should be capable of representing physically the MPS it talks about.\footnote{This is of course not the case for those philosophical positions that take actual observation as their fundament. See for discussion: \cite{deRonde14a}.} 

\begin{definition}
{\bf Meaningful Physical Statements (MPS):} If given a specific situation a theory is capable of predicting in terms of definite physical statements the outcomes of possible measurements, then such physical statements are meaningful relative to the theory and must be constitutive parts of the particular representation of physical reality that the theory provides. Measurement outcomes must be considered only as an exposure of the empirical adequacy (or not) of the theory. 
\end{definition}
 
\noindent It must be remarked that a MPS pertains to the physical representation provided by the particular theory, this means we leave open ---going against Bohr's second desiderata--- the possibility that experience is not reduced to actual measurement outcomes. It should be also noticed that this definition of MPS should be only followed by those who attempt to provide an objective description of physical reality in representational realist terms. Because MPS pertain to physical representation, from a representational realist account of physics, MPS are necessarily related to counterfactual reasoning. 

\begin{definition}
{\bf Counterfactual Reasoning (CR):} The possibility to make MPS in terms of a physical representation allows in general for counterfactual statements in physical discourse. If the theory is empirically adequate then such MPS are taken to be that of which the theory talks about. MPS are not necessarily statements about future events, they can be also statements about past and present events. CR about MPS comprise all actual and non-actual physical experience. 
\end{definition}

If we accept the RRPT stance (section 1), physical representation must take into account the MPS produced by the theory. This also implies that CR is a {\it necessary condition} that a theory must uphold for without it there is no possibility of physical representation nor physical discourse. Without CR in physical discourse one cannot imagine experience beyond actuality. For a representational realist, the power of physics is CR itself, it is the capability that allows us to predict that ``if I perform this or that experiment'' then ---if it is a MPS--- the physical theory will tell me that ``the outcome will be $x$ or $y$'', and I do not need to actually perform the experiment! I know what the result will be independently of actually performing the experiment or not.\footnote{The fact that what we know in QM is given in statistical terms does not imply that we need to explain such predictions in classical terms. Or that knowledge boils down only to `knowledge that accounts for certainty about actuality'. In the case of QM we have a new kind of `certain knowledge in statistical terms'.} That is the whole point of being a realist about physics, that I trust the theory to be talking about a physical representation of reality. CR in physical discourse has nothing to do with time, evolution nor dynamics, it has to do with the possibility of representing experience. A physical theory allows me to make claims about the future, the present and the past, just in the same way physical invariance in classical mechanics connects the multiple frames of reference without me being in any particular one. CR is the discursive invariance with respect to physical phenomena. We do not need to be in a specific frame to know what will happen in that specific frame or a different one. Notice, once again, that CR is a necessary condition only for representational realist approaches, not for those philosophical positions which denying the theory ladenness of physical experience are grounded on raw experimental data.\\

\noindent {\bf Remark:} {\it CR is a necessary condition for a coherent discourse about MPS that pertain to a particular physical representation ---consequently, also for supporting RRPT.}

\section{Getting Rid of the Ghost of Schr\"odinger's Cat} 

``Getting rid of the ghost of Schrodinger's cat'' is a phrase used by Griffiths \cite{Griffiths13} which we find very appropriate to describe the orthodox perspective towards quantum superpositions in the foundational literature. The general uncomfortable feeling with respect to superpositions is always stressed by adjectives like ``embarrassing'', ``weird'', ``strange'', ``spooky'', etc. which always accompany their definition. The truth is that Schr\"odinger's cat is indeed a strange kind of zombie-cat dead and alive at the same time. A creature difficult to picture or imagine. This might be the reason why most interpretations have attempted, either to get rid of what they consider to be a ``ghost'' or simply argue that in ``classical reality'' ghosts do not exist. 

The scope of this paper is not to discuss the many interpretations of QM but rather to analyze their specific viewpoint regarding quantum superpositions. In this respect, we can separate the many interpretations of QM in two main groups according to such understanding. In the first subsection we consider a group of interpretations which avoid discussing about quantum superpositions ``right from the start'' by arguing that QM must be described in terms of ``classical reality''. In the second subsection, we concentrate on a different group which even though (more or less) agrees that QM does not need to be represented in classical terms still argues against the existence of quantum superpositions. Among the many interpretations that follow this second path in this occasion we concentrate in Dieks' modal interpretation and in Griffiths' Consistent Histories (CH) interpretation. According to our analysis, Dieks and Griffiths, although take as a standpoint a realist stance with respect to the interpretation of QM, instead of discussing the possible physical representations of quantum superpositions end up trying to justify measurement outcomes in an instrumentalist fashion. Let us explain this in detail. 

\subsection{Recovering ``Classical Reality''}

As we have argued in \cite{deRonde10} one of the main lines of research ---following the so called Hidden Variable Program (HVP)--- attempts to ``restore a classical way of thinking about {\it what there is}'' \cite[p. 74]{Bacciagaluppi96} through a reconsideration or extension of the formalism. The best known hidden variable theory is due to Bohm's proposal which seems to restore the possibility of discussing in terms of a ASA. In Bohmian mechanics the state of a system is given by the wave function $\psi$ together with the configuration of particles $X$. The quantum wave function must be understood in analogy to a classical field that moves the particles in accordance with the following functional relation: $\frac{dx}{dt} = \nabla S$, where $S = \hbar \delta$ ($\delta$ being the phase of $\psi$). Thus, particles always have a well defined position together with the rest of their properties and the evolution depends on the quantum field. It then follows that, there are no superpositions of states, the superposition is given only at the level of the field and remains as mysterious as the superposition of classical fields. Given a quantum field $\varphi(x)$ the particle will move according to it. If we change the quantum field by adding another filed $\phi(x)$ such that the new quantum field is now the superposition: $\varphi(x)+\phi(x)$, there is no ontological peculiarity involved for now the particle also has a well defined position and will evolve according to the new field. Presumably, due to the fact that the new field is different from the original one the particle will move in a different way and will follow a different trajectory compared to the first case. The field does not only have a dynamical character but also determines the epistemic probability of the configuration of particles {\it via} the usual Born rule.

Apart from the HVP we can also find in the literature the Dynamical Reduction Program (DRP) which, according to Ghirardi \cite{Ghirardi11}, ``consists in accepting that the dynamical equation of the standard theory should be modified by the addition of stochastic and nonlinear terms.'' The most famous of these attempts is the Ghirardi, Rimini and Weber proposal (also called `GRW theory') \cite{GRW} which introduces non-linear terms in the Schr\"odinger equation. The introduction of such non-linear terms attempts to explain the projection postulate and the MP. However, as Ghirardi \cite{Ghirardi11} acknowledges, ``the problem of building satisfactory relativistic generalizations of these models has encountered serious mathematical difficulties due to the appearance of intractable divergences.'' As he makes the point: 

\begin{quotation}
\noindent {\small ``the validity of the superposition principle and the related phenomenon of entanglement [...] have embarrassing consequences since they imply [...] the non-epistemic nature of quantum probabilities, the objective indefiniteness of physical properties and the objective entanglement between spatially separated and non-interacting constituents of a composite system.[...] If one wishes to have an acceptable final situation, one mirroring the fact that we have definite perceptions, one is arguably compelled to break the linearity of the theory at an appropriate stage. [The GRW theory] allows one to understand how one can choose the parameters in such a way that the quantum predictions for microscopic systems remain fully valid while the embarrassing macroscopic superpositions in measurement-like situations are suppressed in very short times.'' [{\it Op. cit.}]}\end{quotation}

A different proposal is that of Many Worlds (MW) interpretation, considered to be a direct conclusion from Everett's first proposal in terms of `relative states' \cite{Everett57}. Staying away from the HVP, Everett's idea was to let QM find its own interpretation, making justice to the symmetries inherent in the Hilbert space formalism in a simple and convincing way \cite{DeWittGraham}. Contrary to the DRP, MW interpretations are no-collapse interpretations which respect the orthodox formulation of QM. The main idea behind this interpretation is that superpositions relate to collections of worlds, in each of which exactly one value of an observable, which corresponds to one of the terms in the superposition, is realized. Apart from being simple, the claim is that it possesses a natural fit to the formalism, respecting its symmetries. The solution proposed by MW to the measurement problem is provided by assuming that each one of the terms in the superposition is {\it actual} in its own correspondent world. According to Everett \cite[p. 146]{Everett73} himself: ``The whole issue of the transition from `possible' to `actual' is taken care of in the theory in a very simple way ---there is no such transition, nor is such a transition necessary for the theory to be in accord with our experience. From the viewpoint of the theory all elements of a superposition (all `branches') are `actual', none any more `real' than the rest.'' Thus, it is not only the single value which we see in `our world' which gets actualized but rather, that a branching of worlds takes place in every measurement, giving rise to a multiplicity of actual worlds with their corresponding actual values. The possible splits of the worlds are determined by the laws of QM but each world becomes again ``classical''. Quantum superpositions are interpreted as expressing the existence of multiple worlds, each of which exists in (its own) actuality. However, there are no superpositions in this, our actual world, for each world becomes again a ``classical world''. The MW interpretation seems to be able to recover these islands of classicality at the price of multiplying the `actual realm'. In this case, the quantum superposition is expelled from each actual world and recovered only in terms of the relation between the multiple worlds.

\subsection{Physical Representation or Measurement Outcomes?}

In this section we concentrate in the arguments provided by Dieks and Griffiths against the existence of quantum superpositions and show how both approaches, going against their original realist stance, end up considering measurement outcomes as the very fundament of their own interpretations. 

First of all we should remark that we agree with Dieks  \cite[p. 189]{Dieks88a} when he argues that: ``It is the state vector which is in a superposition, not the cat itself. `State vector' and `cat' are two concepts at different levels of discourse.'' However, we should also add that the distinction between a mathematical level and a conceptual level happens of course in every physical theory. For example, `a point in phase space' is not the same as `a physical object is space-time'; but the relation between these two levels ---mathematical and conceptual--- is articulated through classical mechanics in such a way that there is in fact a direct relation between `a point in phase space' and `a physical object is space-time'. Indeed, every physical theory ---at least in representational realist terms (section 1)--- has a mathematical formalism which needs to be related to a conceptual network of physical concepts. Every physical notion has a mathematical counterpart, the notion of field, of wave, of particle, they all relate to a mathematical formalism and to specific equations that also constitute and constrain such physical notions. In physics, as we stressed above, it is this interplay between mathematical formalisms and physical concepts which allow us to represent physical experience and reality. And of course this involves a difficult relationship. As Einstein\footnote{Against the interpretation of Einstein as a ``naive'' realist see \cite{Howard93}.} \cite[p. 1196]{PS} clearly made the point:  ``The problem is that physics is a kind of metaphysics; physics describes `reality'. But we do not know what `reality' is. We know it only through physical description...'' The scope of this paper does not intend to discuss the difficult relation between representation and reality, however we take for granted that a serious analysis should leave aside any type of ``naive realism'' which simply neglects the fundamental relevance of representation within the definition itself of phenomena (see for discussion \cite{deRonde14a}). But we should be also careful when analyzing Dieks' statement because the notion of `cat' is a classical notion which represents a classical entity\footnote{A classical entity such as for example a cat is constrained by the logical and ontological principles of existence, non-contradiction and identity. Such principles constrain not only the notion of physical object itself but also its relation to experience.} and it should never be confused with `the cat that is on my table {\it hic et nunc} while I write this paper', for that would be falling into the trap of naive realism; assuming that we have access to reality {\it as it is} by confusing representation with reality.  Physical notions ---even that of  a `cat'--- always pertain to physical discourse and representation, not to reality {\it as it is}. Schr\"odinger designed his 1935 {\it gedankenexperiment} as an {\it ad absurdum} proof which would make explicit how erroneous would be to picture or interpret `quantum superpositions' in terms of classical objects (e.g., a cat). What Schr\"odinger exposed is that a quantum superposition (in the mathematical level) cannot be represented in terms of a classical object (in the conceptual level). Going against the second of Bohr's desiderata (section 2), we believe that the lesson provided by Schr\"odinger should be regarded as an important road-sign that points in the direction of the need of developing new physical concepts that coherently relate to the quantum formalism. 

Dieks himself explains very clearly one of the arguments in favor of the existence of quantum superpositions: 

\begin{quotation}
\noindent {\small ``In classical physics the most fundamental description of a physical
system (a point in phase space) reflects only the actual, and nothing that is merely possible. It is true that sometimes states involving probabilities occur in classical physics: think of the probability distributions $\rho$ in statistical mechanics. But the occurrence of possibilities in such cases merely reflects {\it our ignorance} about what is actual. The statistical states do not correspond to features of the actual system (unlike the case of the
quantum mechanical superpositions), but quantify our lack of knowledge of those actual features. This relates to the essential point of difference between quantum mechanics and classical mechanics that we have already noted: in quantum mechanics the possibilities contained in the superposition state may interfere with each other. There is nothing comparable in classical physics. In statistical mechanics the possibilities contained in $\rho$ evolve separately from each other and do not have any mutual influence. Only one of these possibilities corresponds to the actual situation. The above (putative) argument for the reality of modalities can therefore not be repeated for the case of classical
physics." \cite[pp. 124-125]{Dieks10}}\end{quotation}

\noindent Going against this interpretation, he provides several arguments against the idea ``that there is something real corresponding to each of these individual terms''. Dieks starts his argumentation by stressing that the state $\Psi$ is defined as an element of a mathematical space. ``It does therefore not make immediate sense to speak about causal interactions between them: only physical systems can causally affect each other, whereas numbers, functions or mathematical entities in general, do not have causal effects.'' This first remark also applies of course to any physical theory. Mathematics with no physical interpretation can be only understood as a formal scheme with no relation whatsoever to the physical world (see for discussion \cite{deRondeMassri14}). But as a matter of fact, going back to our earlier example, the evolution of `a point in phase space' in classical mechanics is in fact analyzed in terms of causality. We agree with Dieks that one should be aware of the limits of the mathematical and conceptual levels of discourse, however one cannot deny their tight relation within physical theories. That is the whole point of calling a mathematical equation an ``evolution equation'' or ``equation of motion''. From a realist perspective, we need to provide a physical interpretation of $\Psi$ in order to discuss about causality and physical states of affairs. Contrary to what Dieks claims, this is not an argument against the idea that the terms of a quantum superposition cannot be understood in terms of physical reality but rather the very condition of possibility for such discussion and analysis. As we have argued in the first section of this paper, realism attempts to provide every physical theory with a representation of reality. That is what realism is about. 

Dieks [{\it Op. cit}, p. 131] continues to argue that: ``from a Humean viewpoint it is clear from the outset that the states can only refer to {\it actual} situations.''\footnote{Here actuality seems to be understood as a {\it hic et nunc} actual observation and not in terms of a mode of existence.} We consider this as a correct statement which only shows the problems of empiricism to deal with quantum superpositions. We agree with Dieks that the empiricist perspective has no need to provide a physical representation of quantum superpositions, however, it has to deal instead with two very deep problems. The first deals with the theory ladenness of physical observation, something that could be condensed in the famous dictum that Einstein made to Heisenberg ``It is only the theory which can tell you what can be observed''. Physical observation always presupposes a physical representation. Actual situations need to be described in terms of physical theories, no matter how basic the situation is. As it is well known, the 20th century positivist project of distinguishing between {\it theoretical terms} and {\it empirical terms} failed to provide an understanding of such distinction which is necessary in order to argue in favor of the idea that empirical data is not theory laden (see for discussion \cite{deRonde14a}). The second important point is that from an empiricist perspective there would seem to be no problem whatsoever with QM. For all we know, the theory is in fact empirically adequate, so for an empiricist it seems that should be the end of the road. There is no need for an empiricist to discuss the interpretation of the theory in terms of a representation of physical reality. After all, that is a realist project, not an empiricist one. 

A third point made by Dieks is:

\begin{quotation}
\noindent {\small ``[...] that there is no ground whatsoever to suppose that the plus-sign in our superposition equations stands for simultaneous physical existence [...] that $\Psi = \Psi_1 + \Psi_2$ means that in the situation described by $\Psi$ the situations described by $\Psi_1$ and $\Psi_2$ are physically present as well. In fact, making this assumption would lead to a boundless multiplication of realities, in view of the fact that $\Psi$ can be written as the sum of two other states in an uncountable infinity of ways.'' \cite[pp. 124-125]{Dieks10}}\end{quotation} 

\noindent There are two elements in the argumentation made by Dieks. Firstly he argues against the idea that ``the plus-sign in our superposition equations stands for simultaneous physical existence''. Let us provide a short analysis which shows that there is indeed a physical basis which supports the idea that the ``+'' sign relates to simultaneous physical existence. If we consider a typical Stern-Gerlach type experiment in the $x$-direction we can have the following quantum superposition: $c_{x1} | \uparrow_x\rangle +  c_{x2} | \downarrow_x\rangle$. There are two MPS which relate to each one of the terms in this quantum superposition. Such terms ``evolve'' and ``interact'' according to the Schr\"odinger equation of motion and can be predicted through the Born rule, but since the ignorance interpretation is precluded, one cannot claim that all terms are actual simultaneously. However, this does not mean that all interpretations in terms of simultaneous existence are precluded, it only means that ---by definition of actuality \cite{deRonde15c}--- both terms cannot be considered as simultaneously actual. Indeed, if we assume the metaphysical principle that `Actuality = Reality' it seems difficult to come up with a different interpretation to that of MW. Yet, this can be also understood as a limit imposed by the quantum formalism to the representation of quantum reality in terms of actuality. This is a constrain that leaves open the possibility to develop a non-classical physical representation of reality. We know of no reasonable (non-dogmatic) argument against this possibility. The second element in the argumentation of Dieks is the assumption that ``$\Psi$ can be written as the sum of two other states in an uncountable infinity of ways [...] would lead to a boundless multiplication of realities''. First of all we must remark that Dieks now shifts the question using the fact that $\Psi$ can be represented in terms of different bases. Contextuality enters the scene. Indeed, given our previous SG example, changing the direction of the SG apparatus we can also obtain the quantum superpositions related to the $y$-direction, $c_{y1} | \uparrow_y\rangle +  c_{y2} | \downarrow_y\rangle$, to the $z$-direction, $c_{z1} | \uparrow_z\rangle +  c_{z2} | \downarrow_z\rangle$, and to all possible directions that we can think of. But the important point which supports the idea that each one of these superpositions is related to physical reality is the fact that each one of these superpositions gives rise through the Born rule to a definite set of MPS. This does not seem a strange metaphysical move that multiplies reality beyond necessity, but rather the very condition of empirical adequacy that any interpretation of QM should meet. The numbers that accompany each one of the terms of the different quantum superpositions directly relate (in square modulus) to predictions about experimental outcomes ---$|c_{y1}|^2$ tells us the ratio of finding a result related to  $| \uparrow_y\rangle$, $|c_{y2}|^2$ tells us the ratio of finding a result related to  $| \downarrow_y\rangle$, etc. Such terms ``evolve'' and ``interact'' according to the Schr\"odinger equation of motion. So if every term ``evolves'', ``interacts'' and relates directly to physical experience, turning Dieks' questioning upside down, why shouldn't they be considered as necessary elements in the definition of what the theory is talking about in terms of physical reality?   

Dieks modal interpretation \cite{Dieks88a, Dieks88b, Dieks89a,Dieks05} states the idea that given a quantum state $\Psi$ there is always a basis in which we can mathematically represent $\Psi$ as a superposition of one single term, $|\alpha \rangle$. This of course is not the most general way of mathematically representing $\Psi$, it is in fact the only mathematical representation of $\Psi$ in which the basis has been carefully chosen so that  the number that accompanies the state is 1. In general, $\Psi$  will be represented by a superposition of more than one term, $\sum c_i |\beta_i\rangle$. Dieks continues to argue that the $\Psi$ describes an ASA, namely, the property given by $| \alpha \rangle \langle \alpha |$. However, what Dieks interpretation fails to account for is the physical meaning of all the other quantum superpositions ---with more than one term--- that also give rise to MPS. To make our argument more concrete, and going back to our SG example, given the fact that the SG is placed in the $z$-direction and given the superposition is $| \uparrow_z\rangle$, Dieks modal interpretation claims that physical reality is (only) composed by the property related to the projection operator $| \uparrow_z\rangle \langle \uparrow_z |$. Indeed, if we measure $\Psi$ in the $z$-direction the Born rule predicts that we will obtain with certainty the result related to this property because $|1|^2 = 1$.\footnote{This is analogous to orthodox quantum logic interpretation.} However, this interpretation is unable to provide an account of what really matters for quantum technology, that is, all other superpositions $\sum c_{r_{i}} | \alpha_{r_{i}}\rangle$ (where $r_{i}$ is a particular direction different to $z$ and with $c_{r_i} \neq 0$) that also seem to be part of physical reality. These superpositions with more than one term also provide definite information in terms of MPS about the (non-actual) state of affairs (section 4). Consequently, the description provided by $| \uparrow_z\rangle$ is ---at least--- incomplete. Furthermore, it implicitly changes the subject of inquiry. When I ask, given a SG in the $i$-direction what is the meaning of a quantum superposition given by: $c_{i1} | \uparrow_i\rangle + c_{i2} | \downarrow_i\rangle$  (with $c_{i1} \neq 0$ and $c_{i2} \neq 0$) which provides access to MPS in that specific physical situation, it is not enough to answer ---using the orthodox interpretation of quantum superpositions discussed above--- that, if I change the direction of the SG to the $j$-direction (were I can write $\Psi$ in terms of one single term $| \phi_j\rangle$), I will obtain with certainty the property related to the state $| \phi_j\rangle$. This also evades an answer to the MP (section 3). When I ask about observables related to $| \uparrow_i\rangle$ and $| \downarrow_i\rangle$ ---of which QM provides clear information that relates to physical experience--- the answer cannot be about a different observable related to $| \phi_j\rangle$. The interpretation of Dieks is not capable of providing an answer to the question and thus also to the justification ---in realist terms--- of present technological and experimental developments which  explicitly use Schr\"odinger cats. To put it in a different form, if given a Schr\"odinger cat I ask the question: ``Is the cat `dead' or is it `alive'?'' The answer cannot be: ``The cat is under the table''. 

Since Dieks attempts to recover a description in terms of actualities, he also argues against the reification of modalities: 

\begin{quotation}
\noindent {\small ``I think it is unclear how a realist interpretation of [a probability] $p$ as some kind of ontologically objective chance can help our understanding of what is going on in nature. Clearly, such an interpretation cannot change the empirical content and predictive power of the theory that is involved. But also with regard to explanations nothing seems to be gained by introducing ontologically real chance or real modalities, because the notion of a real modality is in need of explanation itself. For example, we do not really know what kind of things dispositions are, and it is obscure exactly how a disposition could take care of the task of arranging for the right relative frequencies to occur in long series of experiments. Indeed, the very content of the notion of disposition does not seem to go beyond ``something responsible for the actual relative frequencies found in experiments''.''  \cite[p. 133]{Dieks10}}\end{quotation}

\noindent Here we also agree with Dieks that dispositions and propensity type interpretations do not provide any type of addition to the understanding of QM. But our criticism to such interpretations ---which we have presented already in \cite{deRonde11}--- stresses different aspects. Such type of interpretations do not advance truly in the development of a realm of reality independent of actuality ---respecting the classical dogma that: `Actuality = Reality'--- and thus are not able to explain what this new (non-actual) realm is about. Instead, such interpretations define propensities, dispositions and even potentialities as teleological notions in direct relation to actuality and actualization. A `propensity', `dispositional' or `potential' property is one that can become actual.\footnote{We have analyzed in detail such interpretations in \cite[chapter 14]{deRonde11}.}  We believe, contrary to Dieks, that the development of a mode of existence independent of actuality which would match the formalism would provide a new explanatory power in order to think about quantum experience in a radical non-classical manner. If we could think about QM beyond measurement outcomes it is quite possible that we could also understand what the theory is talking about and this, in turn, would also allow us to imagine quantum experience beyond the algorithmic recipe that is being used today.\\ 

Griffiths also discusses the meaning of quantum superpositions trying to ``get rid of the ghost of Schr\"odinger's cat''. His analysis founded on the CH interpretation is very similar to that of Dieks modal interpretation, but in some aspects he makes even more explicit the interpretational maneuvers.  Let us leave to Griffiths the explanation of his own approach:

\begin{quotation}
\noindent {\small ``The CH strategy is to first identify the quantum properties, which is to say the subspaces of the Hilbert space, that should enter the description. Here we are interested in a measurement with two possible outcomes: either detector $a$ has triggered or $b$ has triggered. While the individual states $|D^{a*} \rangle \otimes |D^{b}\rangle $ and $|D^{a} \rangle \otimes |D^{b*}\rangle$ can be interpreted as possible outcomes, their macroscopic superposition $|\Psi_3\rangle$ in (6) [$|\Psi_3\rangle = \alpha |D^{a*} \rangle \otimes |D^{b} + \beta |D^{a} \rangle \otimes |D^{b*}\rangle$] cannot, and indeed the projector [$\Psi_3$] does not commute with any of the projectors [$D^{a*}$], [$D^{a}$], [$D^{b*}$] or [$D^{b}$]. (Our notation follows the usual physicistÕs convention that a projector $P \otimes I$ on a tensor product can be denoted by $P$.) Consequently, if we demand that [$\Psi_3$]  be a physical property at time $t_3$, this choice of framework (i.e.,  [$\Psi_3$]  and $I -  [\Psi_3] $) will prevent us from discussing the situation in Fig. 1 as a measurement, a physical process with some specific macroscopic outcome. Instead we must use a framework at time $t_3$ that contains the projectors [$D^{a}$], [$D^{a*}$], [$D^{b}$] or [$D^{b*}$] and their products. Having made this choice CH employs $|\Psi_3\rangle$ not as a physical property but as a pre-probability, a mathematical device which can be used to calculate the probability of various properties via the usual Born formula: $Pr(D^{a*}) = \langle \Psi_3| D^{a*} \otimes I |\Psi \rangle =  |\alpha|^2$, $Pr(D^{b*}) = |\beta|^2$, $Pr(D^{a*}, D^{b*}) = 0$. (8) (Here $Pr(A, B)$ is the probability of the conjunction $A$ AND $B$.) In words, the probability is $|\alpha|^2$ that the $a$ detector has triggered, $|\beta|^2$ that the $b$ detector has triggered, and 0 that both have triggered. This way of understanding $|\Psi_3\rangle$  is not a new innovation, as it goes back to the work of Born in 1926.'' \cite{Griffiths15}}\end{quotation}

\noindent As we see, Griffiths (like Dieks) distinguishes between superpositions which are written as one single term and those which have more than one term ---using the equivalent class given by equation (6): $|\Psi_3\rangle = \alpha |D^{a*} \rangle \otimes |D^{b} + \beta |D^{a} \rangle \otimes |D^{b*}\rangle$. As Griffiths makes the point:

\begin{quotation}
\noindent {\small ``[...] in the CH discussion of measurement outcomes $|\Psi_3\rangle$ is not regarded as a quantum property, but instead as a pre-probability, a mathematical device employed to calculate probabilities, and hence no more ``real'' than a probability distribution of the sort one encounters in classical statistical mechanics.''  \cite{Griffiths15}}\end{quotation}

\noindent According to Griffiths, when making reference to a superposition of one term one can say, via the Born rule, that the property is actually preexistent; but when we want to claim something about $ \alpha |D^{a*} \rangle \otimes |D^{b} \rangle + \beta |D^{a} \rangle \otimes |D^{b*}\rangle$ Griffiths oblige us to suddenly change the interpretation and think of such superposition in terms of a ``mathematical device employed to calculate probabilities''. But this second interpretation introduced to account for the most general superpositions of more than one term seems to us in no way different to the explicit instrumentalist perspective assumed by Fuchs and Peres when they argue that QM is ``an algorithm for computing probabilities for the macroscopic events  (``detector clicks") that are the consequences of experimental interventions'' (section 1). In a very similar way to Dieks, Griffiths provides a physical representation of only one of the many possible representations of a given $\Psi$, namely, the one in which $\Psi$ is written as one single term. In such case Griffiths argues that ``the property is preexistent'' while in the rest of mathematical representations superpositions are regarded as ``calculational tools'' or algorithmic devices.\footnote{Griffiths also relates the failure to account for quantum superpositions in terms of noncontextuality and the Single Framework Rule in \cite{Griffiths13}. A detailed analysis of this argument is provided in \cite{deRonde15c}.} The fact that probabilities are used does not exempt the CH interpretation of providing a physical meaning to such probability. As a matter of fact, in classical statistical mechanics probability is a very well defined physical notion which describes our ignorance about an ASA. But as Schr\"odinger remarked to Einstein many years ago this is not the case in QM.\footnote{It is well known that QM has a non-Kolmogorovian probability measure which cannot be interpreted in terms of ignorance about an ASA. See for example \cite{Redei12}.}   

\begin{quotation}
\noindent {\small ``It seems to me that the concept of probability is terribly mishandled these days. Probability surely has as its substance a statement as to whether something {\small {\it is}} or {\small {\it is not}} the case ---an uncertain statement, to be sure. But nevertheless it has meaning only if one is indeed convinced that the something in question quite definitely {\small {\it is}} or {\small {\it is not}} the case. A probabilistic assertion presupposes the full reality of its subject.'' \cite[p. 115]{Bub97}}
\end{quotation}

\section{The Ghost in the Lab}

As we have argued above all the interpretations we have discussed seem to remain in line with what we have called `the two Bohrian dogmas' (section 2). None of these interpretations attempts to truly think beyond classicality and its presupposed standpoint that `Actuality = Reality'. Instead, every attempt that we know of ---willingly or not--- ends up trying to build a bridge which can save our ``classical'' actualist understanding of physical reality. And even those approaches which, from a realist standpoint, seem to agree that there is nothing ``classical'' about QM, instead of developing a physical representation of reality that would account for the quantum features in a ``positive'' manner ---as features that should be considered explicitly in order to build up a non-classical representation of physical reality according to the theory--- end up only trying to justify `measurement outcomes'. Luckily enough, quite independently of the discussions in the literature about the MP and the BP which have constantly neglected the physical representation and understanding of quantum superpositions, the work that is being done today in the lab takes as a standpoint the existence of such strange elements of the theory. Indeed, the technical and experimental developments that are taking place today in quantum information processing use as a basic feature of their developments the notion of quantum superposition \cite{Nature5, Nature2, Nature13, Nature1, Nature07, Nature3, Nature4}. Quantum computation, quantum teleportation, quantum cryptography, etc., they all depend on their very existence. If superpositions ---whatever they are--- did not exist, from a representational realist perspective, such developments would not have been possible, it is that simple. 

We agree with Griffiths [{\it Op. cit.}, p. 174] when he argues against: ``the view that the real [quantum] world is classical, contrary to all we have learned from the development of quantum mechanics in the twentieth century.'' However, we must add to this remark that the fact that we can `do things' does not imply that we can `understand them' or `represent them'. This is the main difference, from a realist perspective, between physics and technology, between realism and instrumentalism. Continuing our analysis, on the one hand, quantum superpositions `evolve' and `interact' according to the Schr\"odinger equation of motion (section 5.2), and on the other hand, the outcomes that expose quantum superpositions can be empirically tested through the Born rule. But when in physics a mathematical element of a theory `evolves', `interacts' and can be predicted according to a formalism, then ---always from a representational realist perspective--- that mathematical expression needs to be related (in some way) to physical reality.

\section{Taking Quantum Superpositions (Really) Seriously}

After more than one century of not being able to interpret QM in terms of ``classical reality'', it might be time to admit that QM confronts us with the fact that classical representation of physics is not the end of the road. In this respect a central argument for discussing a realist physical representation of quantum superpositions has to do with the possibility of developing new physical experience. Physical representation allows us to think, and even to imagine new possible physical experiences when a mathematical formalism is coherently supplemented by physical concepts. Physicists are trained to think and imagine situations in terms of the different physical notions which have been created in relation to particular physical theories: waves, particles, fields, etc., allow us to physically represent reality; and it is exactly this possibility, to think beyond measurement outcomes which truly opens the door to new physical experience. This is one of the main reasons why understanding quantum superpositions is so important, for they can become the key to open the door of the quantum realm ---a realm which still today is not understood in coherent realist terms.

\subsection{Quantum Superpositions and MPS}

The quantum wave function $\Psi$ gives rise to clear definite physical statements regarding observables through the Born rule. The MPS provided by QM, statements that have been used in order to develop experimental situations and technological developments, are of the type:  

\begin{definition}
{\bf MPS in QM:} Given a vector in Hilbert space, $\Psi$, the Born rule allows us to predict the average value of (any) observable $O$. 
$$\langle \Psi| O | \Psi \rangle = \langle O  \rangle$$
This prediction is independent of the choice of any particular basis. 
\end{definition}

\noindent This also means that given a $\Psi$, all observables are related to physical reality through MPS independently of the basis. In short, there is no preferred basis according to the formalism.

This definition of MPS in QM implies that according to the formalism all observables ---independently of the context--- must be considered as part of physical (quantum) reality simultaneously and independently of the choice of the context. This is of course ---due to the contextual character of the formalism (see for discussion \cite{deRonde15c})--- not possible if we consider physical reality only in terms of an ASA. This shows that, either the formalism should be changed in order to recover a classical representation of reality or, that we should leave aside our classical representation of physical reality, that is, to consider the logical possibility that `Quantum Physical Reality $\neq$ Actuality'.

\subsection{Quantum Superpositions as Contextual Existents}

The first postulate of QM implies a ``tricky'' definition of what is to be considered a quantum superposition. In any paper we find the following equation defining a superposition of coherent states: 

$$| \psi\rangle =N \ (|a_1\rangle +  \ |a_2\rangle)$$ 

\noindent where $N$ is a normalization factor. This definition is the seed of a deep misunderstanding in the literature which we have analyzed in detail in \cite{deRondeMassri14}. The idea is that $\Psi$ makes reference to something that can be considered ``the same'' independently of the specific representation or context of inquiry. This allows to change the subject of discussion from a superposition $N \ (|a_1\rangle + \ |a_2\rangle)$ of two terms, to a different superposition of only one term $| \psi\rangle$ ---a very particular superposition that can be seemingly interpreted in terms of the actual mode of existence. This interpretation denies the fact that even though $N \ (|a_1\rangle + \ |a_2\rangle)$ and $| \psi\rangle$ are particular mathematical representations of ``the same'' vector in Hilbert space, the equality sign which in strict mathematical terms means a {\it class of equivalences}, should not be confused ---as it is being done today--- with a {\it metaphysical identity}.\footnote{For a discussion regarding this specific point we refer to \cite{daCostadeRonde14, daCostadeRonde15, deRondeMassri14}.} $N \ (|a_1\rangle + \ |a_2\rangle)$ and $| \psi\rangle$ are two different representations of the same vector. To give an example from classical physics, the fact that two different frames of reference in classical mechanics can discuss about the movement of  an object, does not mean that the movement is ``the same'' from different perspectives. The interesting thing to analyze in the equation $N \ (|a_1\rangle + \ |a_2\rangle)= | \psi\rangle$ is not what is ``the same'' but rather what is ``different'' about the two terms in the equation. We should not concentrate so much in the equality sign but in trying to understand in what sense $N \ (|a_1\rangle + \ |a_2\rangle)$ and $| \psi\rangle$ are different. 

It is quite intuitive that each superposition relates to a specific experimental set up, for each supereposition is a representation in terms of a basis (or CSCO). Following this idea we have provided an interpretation of quantum superpositions in terms of the notion of {\it quantum situation} through which one recovers the contextual character of the theory in a natural way \cite{deRonde15a}. Indeed the superposition would then make reference to the specific set of observables considered within a definite set up. The Born rule provides information through the coordinates in square modulus of the rate of definite outcomes that relate to each term in the quantum superposition. Within such physical interpretation it becomes clear that given a $\Psi$, each different superposition will provide a definite set of MPS regarding each set of observables. The interpretation just exposed not only relates the formalism in a more understandable way to the physical experience that is being done today in the lab but also breaks down the orthodox claim that given a $\Psi$ all superpositions are ``the same'' superposition ---a metaphysical claim that is supported by Dieks' and Griffits' interpretations and analysis. Different possibilities of this kind should be investigated.

\subsection{`Contradictory' and `Contrary' Statements about Quantum Superpositions} 

In \cite{daCostadeRonde13}, the author of this paper together with da Costa discussed the possibility to interpret quantum superpositions in terms of a paraconsistent approach. Arenhart and Krause \cite{ArenhartKrause14a, ArenhartKrause14b, ArenhartKrause14c} provided arguments against such Paraconsistent Approach to Quantum Superpositions (PAQS). Some of their arguments where analyzed and criticized in \cite{deRonde15a, deRonde15d}. In this respect  we also call the attention to the ongoing debate regarding the interpretation of quantum superpositions, taking into account the square of opposition, in terms either of `contradictory statements' or `contrary statements'.  We believe that this logical analysis of the formalism can be very useful in order to come up with a coherent interpretation of superpositions.

\subsection{`Measurement' or `Superposition' Problem?}

Finally, we would like to make special emphasis on the fact that if we are willing to truly investigate the physical representation of quantum superpositions then we will need to ``invert'' the MP, meaning that the attention should be focused on the physical representation of the mathematical expression instead of attempting to somehow ``save'' the measurement outcomes. In \cite{deRonde11, deRonde15a} we argued in favor of the possibility of discussing what we call the Superposition Problem (SP). This could be stated in terms of the necessity ---given the new technological era we are witnessing through quantum information processing--- to provide a physical representation of quantum superpositions, leaving aside the MP and its insistence in justifying classical measurement outcomes.\\ 

\noindent {\it {\bf Superposition Problem (SP):} Given a situation in which there is a quantum superposition of more than one term, $\sum c_i \ | \alpha_i \rangle$, and given the fact that each one of the terms relates trough the Born rule to a MPS, the problem is how do we physically represent this mathematical expression, and in particular, the multiple terms?}\\
 
The SP opens the possibility to truly discuss a physical representation of reality which goes beyond the classical representation of physics in terms of an ASA. We are convinced that without such a replacement of the problems addressed in the literature there is no true possibility of discussing an interpretation of QM which provides an objective non-classical physical representation of reality. We know of no reasons to believe that this is not doable.

\section*{Conclusion} 

Quite paradoxically, while experimentalist today are trying by all means to keep Schr\"odinger cats alive in the lab, philosophers of physics have been trying to kill these ghostly creatures in every interpretation of QM that we know of. According to the arguments presented in this paper, quantum superpositions force us, if we are wiling to take a representational realist stance about physical theories, to a radical reconsideration of physical reality itself. Since actuality seems not to be a mode of existence compatible with that of QM, the equation `Actuality = Reality' cannot continue guiding us in the development of a physical representation of reality which matches the formalism of the theory. According to our analysis we should start to seriously consider the possibility that `Quantum Physical Reality $\neq$ Actuality'.

\section*{Acknowledgements} 

We want to thank G. Domenech and N. Sznajderhaus for a careful reading of previous versions of this manuscript. This work was partially supported by the following grants: FWO project G.0405.08 and FWO-research community W0.030.06. CONICET RES. 4541-12 (2013-2014).

\end{document}